\documentstyle[aps,floats,epsfig]{revtex}
\twocolumn
\newcommand{\be}{\begin{equation}}
\newcommand{\ee}{\end{equation}}
\newcommand{\ba}{\begin{eqnarray}}
\newcommand{\ea}{\end{eqnarray}}
\newcommand{\bd}{\begin{displaymath}}
\newcommand{\ed}{\end{displaymath}}
\newcommand{\bi}{\begin{itemize}}
\newcommand{\ei}{\end{itemize}}

\newcommand{\nn}{\nonumber}
\newcommand{\tJ}{$t$-$J$\ }
\newcommand{\tj}{$t$-$J$\ }

\begin{document}
%\wideabs{
\title{A Bosonic Model of Hole Pairs}
\author{Thomas Siller $^{1}$, Matthias Troyer$^{1}$, T.\,M. Rice$^{1}$,
        Steven\,R. White$^{2}$}
\address{
        $1$ Theoretische Physik, 
        Eidgen\"ossische Technische Hochschule, CH-8093 Z\"urich,
        Switzerland \\
        $^2$ Department of Physics and Astronomy, University of California
        Irvine, CA 92697
        }
\date{\today}
\maketitle
\begin{abstract}
We numerically investigate a bosonic representation for hole pairs 
on a two-leg $t$-$J$ ladder where hard core bosons on a chain
represent the hole pairs on the ladder.
The interaction between hole pairs is obtained by fitting the
density profile obtained with the effective model to the one obtained with 
the \tj model, taking into account the inner structure
of the hole pair given by the hole-hole correlation function.
For these interactions we calculate the Luttinger liquid
parameter, which takes the universal value $K_{\rho}=1$ as half filling
is approached, for values of the rung exchange $J'$ between strong coupling and
the isotropic case.
The long distance behavior of the hole-hole correlation function is also
investigated. Starting from large $J'$, the correlation length first
increases as expected, but diminishes significantly as $J'$ is reduced
and bound holes sit mainly on adjacent rungs.
As the isotropic case is approached, the correlation length increases again.
This effect is related to the different kind of bonds in the region between
the two holes of a hole pair when they move apart.
\end{abstract}
%}
%\vskip2pc] \narrowtext
%
%------------------------------------------------------------------------------
% Section I : Introduction
%------------------------------------------------------------------------------
%
\section{Introduction}
The study of strongly correlated electrons in ladder systems
has been an active field in recent years. Ladders consisting
of a number of coupled legs (or chains) show a rich variety 
of phases \cite{balentsfisher,dagotto,sigrist,tsunetsugu,troyer,dagotto_rice}
depending on the number of 
legs and the electron density.
One property of particular interest is the binding of holes
into pairs when weakly doped into a half-filled two leg 
ladder, to form a Luther-Emery liquid.
In this liquid the spin degrees of freedom are gapped and 
the low energy sector can be mapped to a boson liquid of
hole pairs. This can be studied analytically in the weak
coupling limit described, for example, by a one-band Hubbard
model with weak interactions. In the strongly interacting
limit described by a \tj model,\cite{zhangrice} one must use numerical
methods such as Lanczos diagonalization and the recently
developed Density Matrix Renormalization Group (DMRG)
method.\cite{srwhite,peschel}
Large simulations can be carried out using the DMRG
yielding information on the equal time correlations.

In this paper our goal is to use the hole density
distributions determined by DMRG to determine the form of
the effective boson model that we know on general grounds 
describes the hole pairs. We use these DMRG results to 
parameterize their dispersion and their effective mutual
interactions and also the interaction with hard wall boundary
conditions that occur in DMRG.
In this first paper we limit ourselves to the simplest case
of the two leg ladder. The same approach can be made for
wider ladders with an even number of legs and is currently
in progress.

We start with the \tj ladder Hamiltonian 
\ba
\label{tJham}
H &=& -t \sum_{i,j,\sigma} {\cal P}(c^\dag_{i,j,\sigma}  c_{i+1,j,\sigma} +
c^\dag_{i+1,j,\sigma}  c_{i,j,\sigma}){\cal P} \\ \nonumber
&& -t' \sum_{i,\sigma} {\cal P}(c^\dag_{i,1,\sigma}  c_{i,2,\sigma} +
c^\dag_{i,2,\sigma}  c_{i,1,\sigma}){\cal P} \\ \nonumber
&&+ J \sum_{i,j} ({\mathbf S}_{i,j}{\mathbf S}_{i+1,j}
  -\frac{1}{4} n_{i,j}n_{i+1,j}) \\ 
&&+ J' \sum_{i} ({\mathbf S}_{i,1}{\mathbf S}_{i,2}
  -\frac{1}{4} n_{i,1}n_{i,2})\nn
\ea
where $i$ runs over $L$ rungs, $j(=1,2)$ and $\sigma$
$(=\uparrow,\downarrow)$ are leg and spin indices. The projection operator
${\cal P} \equiv \prod_{i,j}(1-n_{i,j,\uparrow}n_{i,j,\downarrow})$
prohibits double occupancy of a site. Unless noted otherwise we set 
$t=t'$ and $J=0.35\, t$.\\

%
%------------------------------------------------------------------------------
% Section II : One hole pair
%------------------------------------------------------------------------------
%
\section{The structure of a hole pair}
In this section we examine a hole pair on a \tj ladder, studying
the internal hole-hole correlation (hhc) function. The hhc function
gives us information about the inner structure of a hole pair and
depends on the parameters $t$, $t'$, $J$ and $J'$ in the \tj
Hamiltonian given above.
Since the DMRG computations are performed with open boundaries,
special care must be taken in the measurement of the hhc-function,
which we define as 

\be
g_j(r_1,r_2) =
\frac{\langle n_{r_1,1}^h\,
n_{r_2,j}^h\rangle}{\langle n_{r_1,1}^h\rangle 
\langle n_{r_2,j}^h \rangle}\, n^h \ .
\ee
The index $j=1,2$ denotes intra- and inter-leg-correlations
respectively, $n_{i,j}^h=1-n_{i,j}$ denotes the hole density
at site $(i,j)$, and $n^h$ the average hole density. Introducing
relative and center of mass coordinates, $r$ and $R_0$, we
measure $g_j(r,R_0)$ on ($40\times 2$)-ladders with open
boundaries for $R_0 \simeq L/2$.
In a range $R_0 = L/2\pm 5$, $g_j(r,R_0)$ depends only weakly
on $R_0$. Furthermore, we have tested that better accuracy for the
hhc-function cannot be achieved on larger ladders up to ($60\times 2$)
sites. We conclude that these measurements give good values for
$g_j(r)$ on an infinite ladder.

\begin{figure}
\begin{center}
\epsfxsize=\linewidth
\epsffile{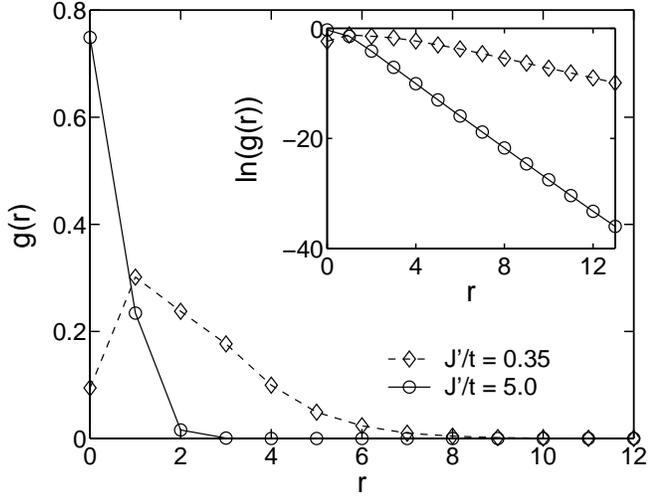}
\end{center}
\caption{Hole-hole-correlation function $g(r)$ obtained from DMRG 
  calculations on a ($40\times 2$) \tj ladder for $J' = 0.35\, t$ 
  and  $J' = 5.0\, t$ with one hole pair ($J = 0.35\, t$, $t = t'$).}
\label{f01}
\end{figure}

We denote by
\be\label{gr}
g(r)=\sum_j g_j(r)
\ee
the sum of intra- and inter-leg correlations.
We have examined $g(r)$ as a function of the rung exchange
coupling $J'$.
As can be seen in Fig.~\ref{f01}, a significant change in the hhc-function
occurs when the isotropic case ($J' \to J$) is approached from the strong
rung coupling regime ($J' \gg J$).
In the latter case, the holes sit mainly on the same rung whereas in the
former case, they are found predominantly on adjacent rungs.
We study the long distance behavior of the hhc-function by
fitting the tail to an exponentially decaying function,
$g(r)\sim \exp(-r/\xi_{hhc})$.
The results are shown in Fig.~\ref{xicorr}, where $\xi_{hhc}$
is plotted as a function of $J'$.
Starting from strong rung coupling, the correlation length first 
increases as expected, but decreases significantly as $J'$ is
reduced and bound holes sit mainly on adjacent rungs. As the isotropic
case is approached, the correlation length increases again.

\begin{figure}
\begin{center}
\epsfxsize=\linewidth
\epsffile{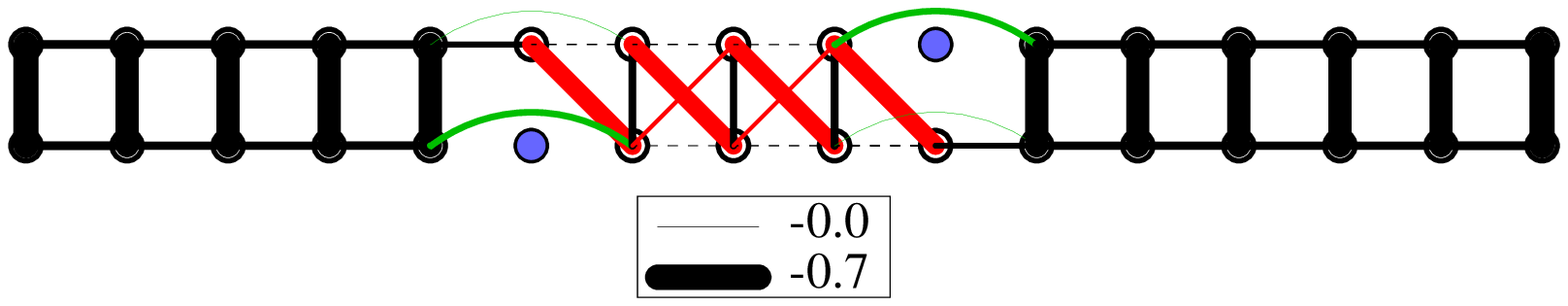}\\
\epsfxsize=\linewidth
\epsffile{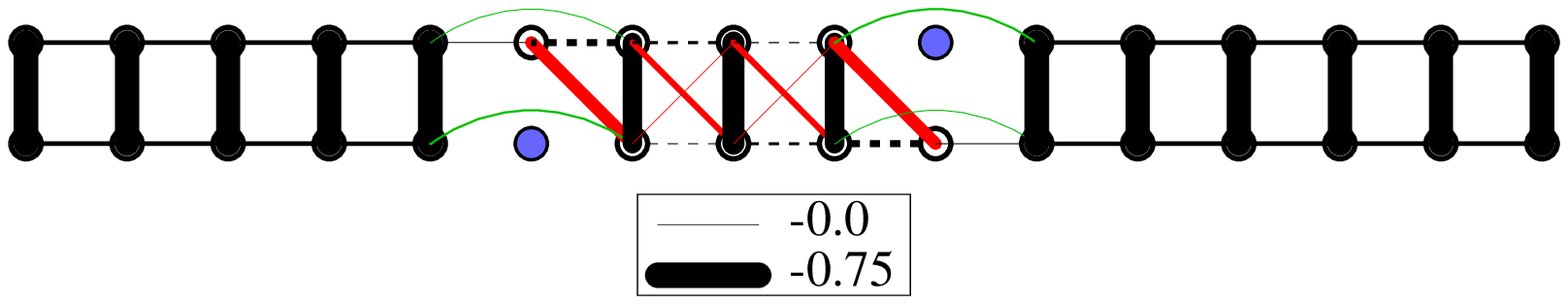}
\caption{
  Exchange bonds $\langle\, {\mathbf S}_i\cdot{\mathbf S}_j\,\rangle$
  around two dynamic holes after projection: $t=t'$, $J=0.35\, t$,
  $J'= 1.2\,t$ for the upper and $J=0.35\, t$, $J'=0.70\, t$ for the
  lower figure. The thickness of the lines is proportional to
  the bond strength.
}
\label{srw01}
\label{srw02}
\end{center}
\end{figure}

This behavior can be explained by looking at the different kinds of bonds
in the region between two holes of a pair as the holes move apart.
In Fig.~\ref{srw01}, we show the exchange fields around a pair of holes.
These are obtained by measuring ${\mathbf S}_i\cdot{\mathbf S}_j$
after projecting out a par\-ticular configuration of two
holes from the ground state $|\psi\rangle$ given for two holes
on a ($16\times 2)$-ladder.
Denoting the corresponding projection operator by $P_h$,
$P_h\, {\mathbf S}_i\cdot{\mathbf S}_j\, P_h$ has been measured and 
normalized by $\langle \psi | P_h |\psi\rangle$.
The procedure is well described by White and Scalapino.
\cite{srw_pair_structure}
This kind of measurement gives us a ``snapshot'' of the spin configuration
around a dynamic hole.
In the following we use the term ``bond'' simply to indicate that
$\langle\, {\mathbf S}_i\cdot{\mathbf S}_j\,\rangle < 0$. 
If this expectation value is close to $-0.75$ for
two sites $i$ and $j$, we say that there is a ``singlet bond'' 
connecting $i$ and $j$.

For $J'> 1.2\, t$ the rung exchange dominates. When a hole pair
virtually splits up, the region between the two holes reverts to rung 
singlets, as shown in Fig.~\ref{srw01}. A high energy is needed 
to split up the holes, but after they are separated there is no further 
increase in energy in moving them apart, since the diagonal exchange 
bonds are weak.
For $J'<0.7\, t$ the region between the holes does something more
complicated, which facilitates hole hopping at the expense of higher 
exchange energy on the intervening rungs.
As shown in Fig.~\ref{srw02}, various diagonal exchange bonds 
are formed, which turn into the rung singlets when the holes hop 
back together. 
This allows for easy hopping and low kinetic energy, but the
exchange energy cost is high. For $J'=0.7\, t$, more and more
distorted exchange bonds are created as the holes move apart.
This leads to a stronger interaction for larger separation than
in the case for $J'= 1.2\, t$.
This weakening of the long range interaction between the two holes
explains the increase of $\xi_{hhc}$ in Fig.~\ref{xicorr}
in the range between $J'=0.7\, t$ and $J'\simeq 1.2\, t$.

Somewhat similar analysis can be found in  
Ref.~\onlinecite{srw_pair_structure}
where the authors have argued that the exchange bonds seen in 
Fig.~\ref{srw02} are responsible for the pairing seen in the isotropic 
case, as well as for stripe formation.

%------------------------------------------------------------------------------
% Section III : The effective Model
%------------------------------------------------------------------------------
%
\section{The effective Model for Hole Pairs}

In the strong coupling limit $J'\gg J$, in the ground state,
two bound holes are on the same rung and a description of
tightly bound hole pairs moving in a background of singlet
rungs is appropriate.
Considering these hole pairs as hard core bosons (hcb) one can
map to an effective boson model as discussed in Ref.~\onlinecite{troyer}
and obtain the hole density directly from the corresponding hcb
density as well as the correlations between hole pairs.

\begin{figure}
\begin{center}
\epsfxsize=\linewidth
\epsffile{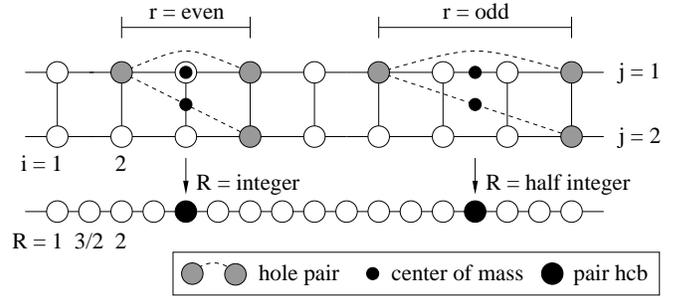}
%\\[3mm]
\caption{
  Mapping of pairs of holes from a ladder to
  hcb on a chain depending on the different positions
  of their center of mass and their relative separation
  $r$ along the legs. R denotes the center of mass
  coordinate along the legs.
}
\label{ladder02}
\end{center}
\end{figure}

Here we propose an effective model that applies for any value of
$J'$ between strong coupling and the isotropic case.
Since holes on a ladder can pair with more weight on adjacent
rungs, our effective model should incorporate the possibility
that the ``center of mass'' of a hole pair can lie on a rung or
between two rungs as shown in Fig.~\ref{ladder02}.
Note that for even (odd) distance $r$ along the legs between
two holes, the center of mass lies on a rung (between two rungs).

To motivate our effective model we study first the case of pairing
between two holes on a two-leg ladder with length L where the
Hamiltonian is given by
\ba\label{model1}
H&=&-t_h \sum_{j=1}^2 \sum_{i=1}^L (b_{i,j}^\dag b_{i+1,j} 
+ b_{i+1,j}^\dag b_{i,j}) \\
&&-t_h \sum_{i=1}^L (b_{i,1}^\dag b_{i,2} 
+ b_{i,2}^\dag b_{i,1}) + V_h \ \ .\nn 
\ea
Here $b^\dag_i$ and $b_i$ denote hole creation and annihilation
operators respectively. Only single occupation of a site
is allowed and we use periodic boundary conditions. 
We choose the attractive interaction $V_h$ between the two holes
strong enough to give an even parity bound state.
We want to find an effective Hamiltonian for the paired holes,
expressed in terms of a new effective boson operator $B_i^{\dag}$
which describes the pair as a whole, and then answer the question,
how do we obtain the hole density $n_{i,j}^b=b_{i,j}^{\dag} b_{i,j}$
from the density $N_i=B_i^{\dag}B_i$ of the effective bosons?

For the Hamiltonian above, with the help of the operators 
\ba
p_{R,r}^{\dag} &=& \frac{1}{\sqrt 2}\,
(b_{R-\frac{r}{2},1}^{\dag} b_{R+\frac{r}{2},1}^{\dag} +
b_{R-\frac{r}{2},2}^{\dag} b_{R+\frac{r}{2},2}^{\dag})\,(1-\delta_{r,0})\\
\tilde p_{R,r}^{\dag} &=&
\left\{
\begin{array}{ll} 
\frac{1}{\sqrt 2} \, (b_{R-\frac{r}{2},1}^{\dag} b_{R+\frac{r}{2},2}^{\dag} +
b_{R-\frac{r}{2},2}^{\dag} b_{R+\frac{r}{2},1}^{\dag}) & 
r \ne 0 \\
b_{R-\frac{r}{2},1}^{\dag} b_{R+\frac{r}{2},2}^{\dag} &
r = 0
\end{array}
\right.
\ea
the wavefunction $\Psi^\dag |0 \rangle$ for the ground state can be written as
\ba
\Psi^\dag &=& 
\frac{1}{\sqrt L} \sum_{R={\rm int.}}\sum_{r_{\rm even}} 
\left(
\varphi_{1}(r) \, p^{\dag}_{R,r} + 
\varphi_{2}(r) \, \tilde p^{\dag}_{R,r}
\right) \\ 
&&
+
\frac{1}{\sqrt L} \sum_{R={\rm half}}\sum_{r_{\rm odd}} 
\left(
\varphi_{1}(r) \, p^{\dag}_{R,r} + 
\varphi_{2}(r) \, \tilde p^{\dag}_{R,r}
\right)\ .\nn
\ea
Here $R$ and $r (=0,1,2,\dots)$ denote the center of mass and relative 
coordinates along the legs as sketched in Fig.~\ref{ladder02}. 
$\varphi_j(r)$ depends only on $r$ and the interaction $V_h$ in 
Eq.~(\ref{model1}).
The pair correlation function $g_j(r)$ for the holes $b^{\dag}_i$ in this
model is simply given by $g_j(r)=|\varphi_j(r)|^2$, where $j=1,2$ denote
intra- and inter-leg correlations respectively.

The probability of finding the center of mass of a pair centered on a rung,
$w_{\rm int}$, or between two rungs, $w_{\rm half}$, is given by
\ba\label{wevenodd}
w_{\rm int}&=&
\sum_{r_{\rm even}} |\varphi_1(r)|^2 + |\varphi_2(r)|^2 = \sum_{r_{\rm even}}
g(r)\\ 
w_{\rm half}&=&
\sum_{r_{\rm odd}} |\varphi_1(r)|^2 + |\varphi_2(r)|^2 = \sum_{r_{\rm odd}}
g(r) \nn
\ea
where $g(r)$ denotes the sum of the inter- and intra-leg
correlations as defined in Eq.~(\ref{gr}). 
Note that the probability to find the pair on or
between rungs depends only on whether $R$ is integer
or half integer.

The same occupation probabilities can be obtained with
the Hamiltonian
\ba\label{model2}
H&=&-t^* \!\!\!\!\! \sum_{R=\frac{1}{2},1,\, \dots}^{L} 
\!\!\!(B_{R}^\dag B_{R+\frac{1}{2}} 
+ B_{R+\frac{1}{2}}^\dag B_{R}) + \epsilon \!\!\! 
\sum_{R=\frac{1}{2},\frac{3}{2},\, \dots}^{L+\frac{1}{2}} \!\!\! N_{R}
\ea
%\ba\label{model2}
%H&=&-t^*\sum_{R=\frac{1}{2},1,\, \dots}^{L} 
%(B_{R}^\dag B_{R+\frac{1}{2}} 
%+ B_{R+\frac{1}{2}}^\dag B_{R}) 
%\\
%&& + \epsilon \sum_{R=\frac{1}{2},1,\, \dots}^{L} \delta_{1,(-1)^{2R+1}}
%N_{R}
%\nonumber
%\ea
for one boson $B_R^\dag$ which moves on a closed chain of length $L$
under the action of a periodically varying onsite potential $\epsilon$.
Here $N_R=B^{\dag}_R B_R$.

Figure~\ref{ladder02} shows the mapping of hole pairs
from the ladder to effective bosons on a single chain.
Note that the center of mass coordinate $R$ of the pairs
determines the position of the effective boson.
Here the ratio of the probabilities $w_{\rm half}/w_{\rm int}$
to find the boson on a site with half integer or integer $R$
depends only on $\epsilon/t^*$ and can easily be obtained as
\be
\frac{w_{\rm half}}{w_{\rm int}} = 
\frac{\epsilon^2+(4t^*)^2-\epsilon \sqrt{\epsilon^2+(4t^*)^2}}
{\epsilon^2+(4t^*)^2+\epsilon \sqrt{\epsilon^2+(4t^*)^2}} \ .
\ee \\
From this equation and Eq.~(\ref{wevenodd}) we obtain for $\epsilon$ 
\be\label{epsilon}
\epsilon=
4\,t^* \sinh{(\ln{\biggl|\sqrt{\frac{\sum_{r_{\rm even}}
g(r)}{\sum_{r_{\rm odd}} g(r)}}\biggr|})} \ .
\ee

The boson operator $B_R^{\dag}$ can be expressed in terms of the
operators $p_{R,r}^{\dag}$ and $\tilde p_{R,r}^{\dag}$ as
\ba \label{pairbosonoperator}
B_R^{\dag}&=&
\left\{
\begin{array}{cl}

\frac{
\sum_{r_{\rm even}} 
\left(
\varphi_{1}(r) \, p^{\dag}_{R,r} + 
\varphi_{2}(r) \, \tilde p^{\dag}_{R,r}
\right)
}
{\left[{\sum_{r_{\rm even}} g(r)}\right] ^{\frac{1}{2}}}
&
\textrm{for int. $R$}
\vspace{.2cm}
\\

\frac{
\sum_{r_{\rm odd}} 
\left(
\varphi_{1}(r) \, p^{\dag}_{R,r} + 
\varphi_{2}(r) \, \tilde p^{\dag}_{R,r}
\right)
}
{\left[{\sum_{r_{\rm odd}} g(r)}\right] ^{\frac{1}{2}}}
&
\textrm{for half int. $R$.}
\end{array}
\right
. 
\ea
Once the density $N_R$ for the model (\ref{model2}) has been computed,
we obtain the density $n_{i,j}^b = b^{\dag}_{i,j} b_{i,j}$ for the model
(\ref{model1}), taking into account the inner structure of the effective
boson (\ref{pairbosonoperator}) by the convolution
\ba\label{convolution}
n_{i,j}^b &=&
\frac{1}{2} \frac{\sum_{r_{\rm even}} g(r) (N_{i-\frac{r}{2}}
 + N_{i+\frac{r}{2}})} {\sum_{r_{\rm even}} g(r)}\\
&&+
\frac{1}{2} \frac{\sum_{r_{\rm odd}} g(r) (N_{i-\frac{r}{2}}
 + N_{i+\frac{r}{2}})} {\sum_{r_{\rm odd}} g(r)} \ .\nn
\ea

We are interested in the ground state properties of the \tj model.
Since the spin part of the ground state wavefunction is antisymmetric
and the spin excitations are gapped,\cite{troyer} the charge degrees
of freedom can be described considering hole pairs as effective 
(hard core) bosons moving in a spin liquid.

To obtain the effective Hamiltonian describing hole pairs
in the \tj model (\ref{tJham}), we model the holes as in
Eq.~(\ref{model1}) and define in analogy to the preceding
\ba
\label{modelHamiltonian}
H &=& -t^* \!\!\! \!\! \sum_{R=1,\frac{1}{2},\,\dots}^{L-\frac{1}{2}}
\!\!\!(B^\dag_R B_{R+\frac{1}{2}} + B^\dag_{R+\frac{1}{2}} B_R)
 \\ \nonumber && 
+ \epsilon \!\!\! \sum_{R=\frac{3}{2},\frac{5}{2},\,\dots}^{L-\frac{1}{2}}
\!\!\!  N_R + V_{\rm int} + V_{b}
\ea
where $N_R = B^\dag_R B_R$ and the $B^\dag_R$ and $B_R$
denote hcb creation and destruction operators respectively.
Since we are using open boundaries, we have to take into
account the interaction of the hole-pairs with the
boundaries.
The potential  $V_{b}$ has been introduced to describe
this effect.
The potential $V_{\rm int}$ gives the interaction between hcb,
i.e. hole pairs in the \tj model.
The onsite potential $\epsilon$ on half-integer sites has
the same meaning as explained above.

%
%------------------------------------------------------------------------------
% Onsite potential
%------------------------------------------------------------------------------
%
\section{Computing the model parameters}

In this section we determine
the effective model parameters $t^*$, $\epsilon$ and
the interactions $V_b$ and $V_{\rm int}$.

\subsection{Onsite potential $\epsilon$}

\begin{figure}
\begin{center}
\epsfxsize=\linewidth
\epsffile{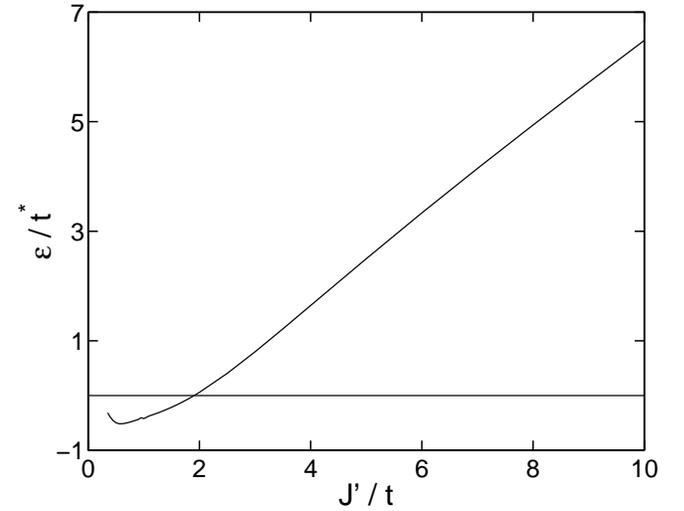}
\end{center}
\caption{The onsite potential $\epsilon$ obtained from Eq.~(\ref{epsilon})
  as a function of $J'$ ($J = 0.35\, t$, $t = t')$.  The hhc-functions were
  obtained from DMRG calculations on ($40\times 2$) \tj ladders.}
\label{f02}
\end{figure}

The onsite potential $\epsilon$ has been calculated with the
hhc-function $g(r)$ obtained numerically by the DMRG for two
holes on ($40\times 2$) \tj ladders for various $J'$. The
results are given in Fig.~\ref{f02}.
They clearly show that for the isotropic case, i.e. $J=J'$, where 
$\epsilon <0$, the hole pair is mainly centered between two rungs,
whereas for strong coupling, i.e. $J'\gg J$, where $\epsilon > 0$,
both holes of a pair sit on the same rung and can be well described
by the effective model given in Ref.~\onlinecite{troyer}.

%
%------------------------------------------------------------------------------
% Interaction with the boundaries
%------------------------------------------------------------------------------
%

\subsection{Interaction with the boundaries}

\begin{figure}
\begin{center}
\epsfxsize=\linewidth
\epsffile{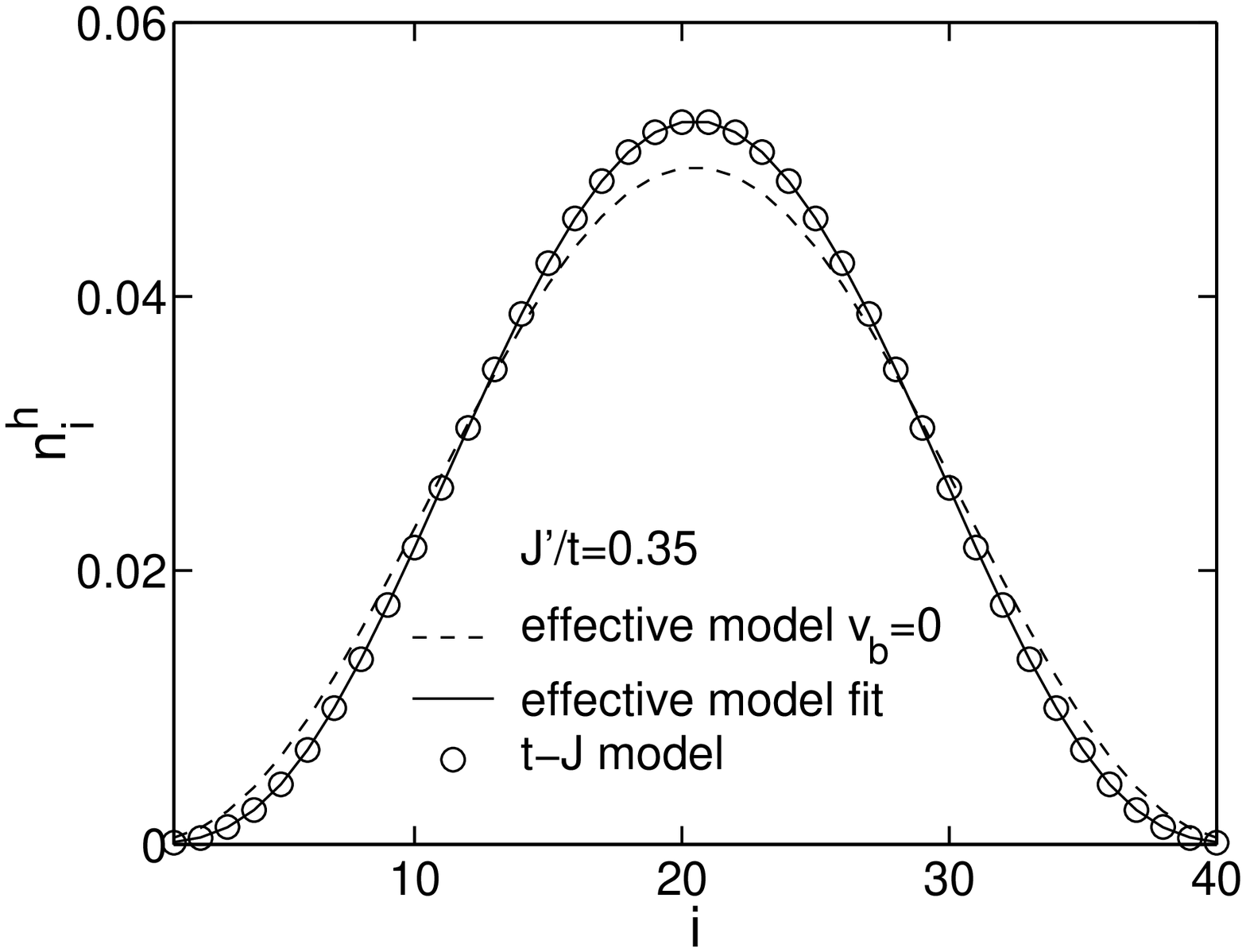}\\
\epsfxsize=\linewidth
\epsffile{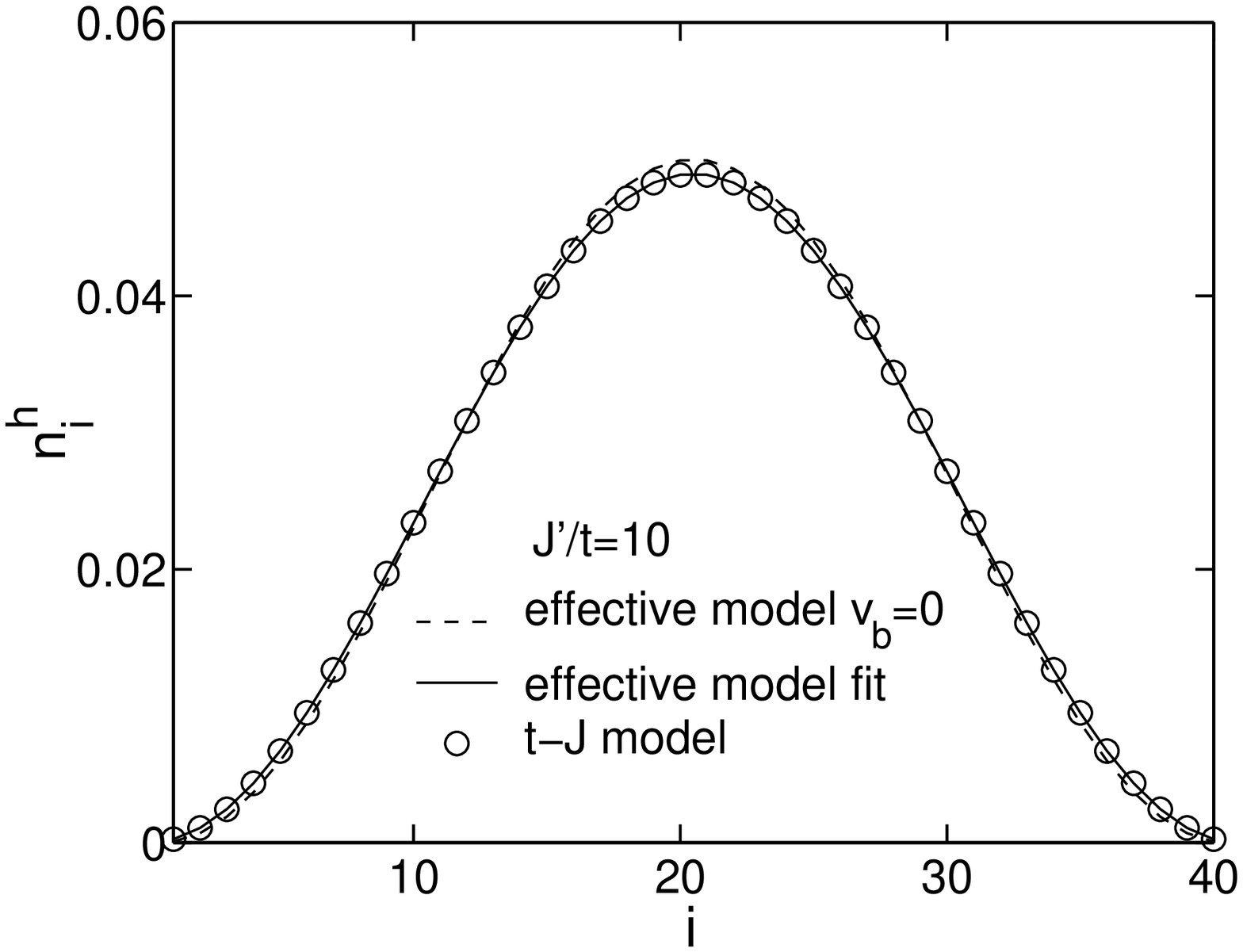}
\caption
{
  Hole density $n_i^h$ for two holes on a ($40\times 2$) \tj ladder with
  $J'=0.35\, t$ and $J'=10\, t$ ($J = 0.35\, t$, $t = t'$) computed
  directly and with the effective model. For the effective model the data
  for $v_b=0$ and the best fit is shown.
} 
\label{edgepotfitfig1}
\end{center}
\end{figure}

\begin{figure}
\begin{center}
\epsfxsize=\linewidth
\epsffile{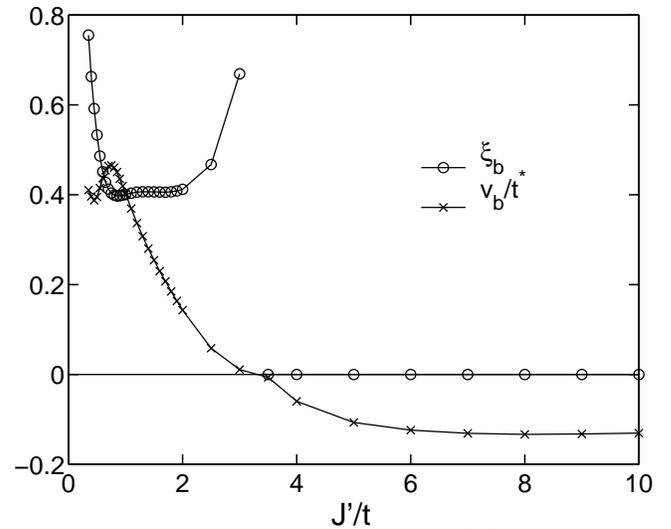}
\caption{
  Parameters for $V_b$ given in Eq.~(\ref{vb}) obtained from fits
  with ($40\times 2$) \tj ladders with one hole pair
  ($J = 0.35\, t$, $t = t'$).
}
\label{edgepotfit}
\end{center}
\end{figure}

As $\epsilon$ is determined by the hhc-function alone, we can compute
the hcb density $N_i$, convolute it with $g(r)$ according to
Eq.~(\ref{convolution}) and compare it with the hole density $n_i^h$ 
of the corresponding \tj system. The hole density $n_{i,j}^h$ can be
obtained using Eq.~(\ref{convolution}) by identifying $n_{i,j}^b$ with
$n_{i,j}^h$.
In this way we obtain $V_{b}$ and $V_{\rm int}$ by fitting the density
profile obtained with the effective model to the one obtained with the
\tj model.
Since $V_{\rm int}$ in Eq.~(\ref{model2}) gives no contribution for one
hcb, we can obtain $V_{b}$ by considering one hole pair in the \tj model.
We choose an exponentially decreasing form for $V_b$
%\ba\label{vb}
%V_{b}&=&v_{b} \sum_{R} N_R \, \left(\exp(-\frac{R-1}{\xi_{b}})+
%\exp(-\frac{L-R}{\xi_{b}})\right)
%\ea
%
\ba\label{vb}
V_{b}&=&v_{b} \sum_{R} N_R \, \left(e^{-\frac{R-1}{\xi_{b}}}+
e^{-\frac{L-R}{\xi_{b}}}\right)
\ea
%
%\ba\label{vb}
%V_{b}&=&2\, v_{b} \sum_{R} N_R \, \cosh(\frac{R-\frac{L+1}{2}}{\xi_{b}})
%\ea
with increasing distance from the boundary.
Figure~\ref{edgepotfitfig1} shows fits for $J'=0.35\,t$ and $J'=10\,t$. and the
parameters $v_{b}$ and $\xi_{b}$ are shown in Fig.~\ref{edgepotfit}.
Up to $J'=3.0\, t$, $v_{b}$ is positive and $\xi_{b}$ is finite.
Above this value $v_{b}$ becomes negative and $\xi_{b}$ is zero.
In this case only the outermost rungs become attractive for hole
pairs as can be expected for large $J'$, due to the charge part
$-\frac{1}{4}\,n_{i,j}^h\,n_{i+1,j}^h$ of the $J'$ term.

%
%------------------------------------------------------------------------------
% Interaction between the hcb (hole pairs)
%------------------------------------------------------------------------------
%

\subsection{The interaction term $V_{\rm int}$}

To obtain the interaction potential $V_{\rm int}$ we proceed in the same
way as for $V_b$. We choose a hard core analytical form
\ba\label{interaction}
V_{\rm int} &=& \sum_R \sum_{R'>R} v_{{\rm int}}(|R-R'|)\, N_R N_{R'} \\
\nonumber
v_{{\rm int}}(r) &=& \left\{ 
\begin{array}{ll}
\infty & \textrm{if $r<r_{min}$} \\
v_1 & r=r_{min}\\
v\, e^{-\frac{r-r_{min}-\frac{1}{2}}{\xi}} & \textrm{if $r > r_{min}$}
\end{array}
\right.
\ea
with $r_{min} \ge 1$ and considered ($30\times 2$) \tj ladders with
four holes and the corresponding effective model with two hcb.
The infinite repulsion for $r=\frac{1}{2}$ reflects the hard core 
condition for hole pairs, so that the minimum distance of two hole
pairs in the \tj model is 1.

\begin{figure}
\begin{center}
\epsfxsize=\linewidth
\epsffile{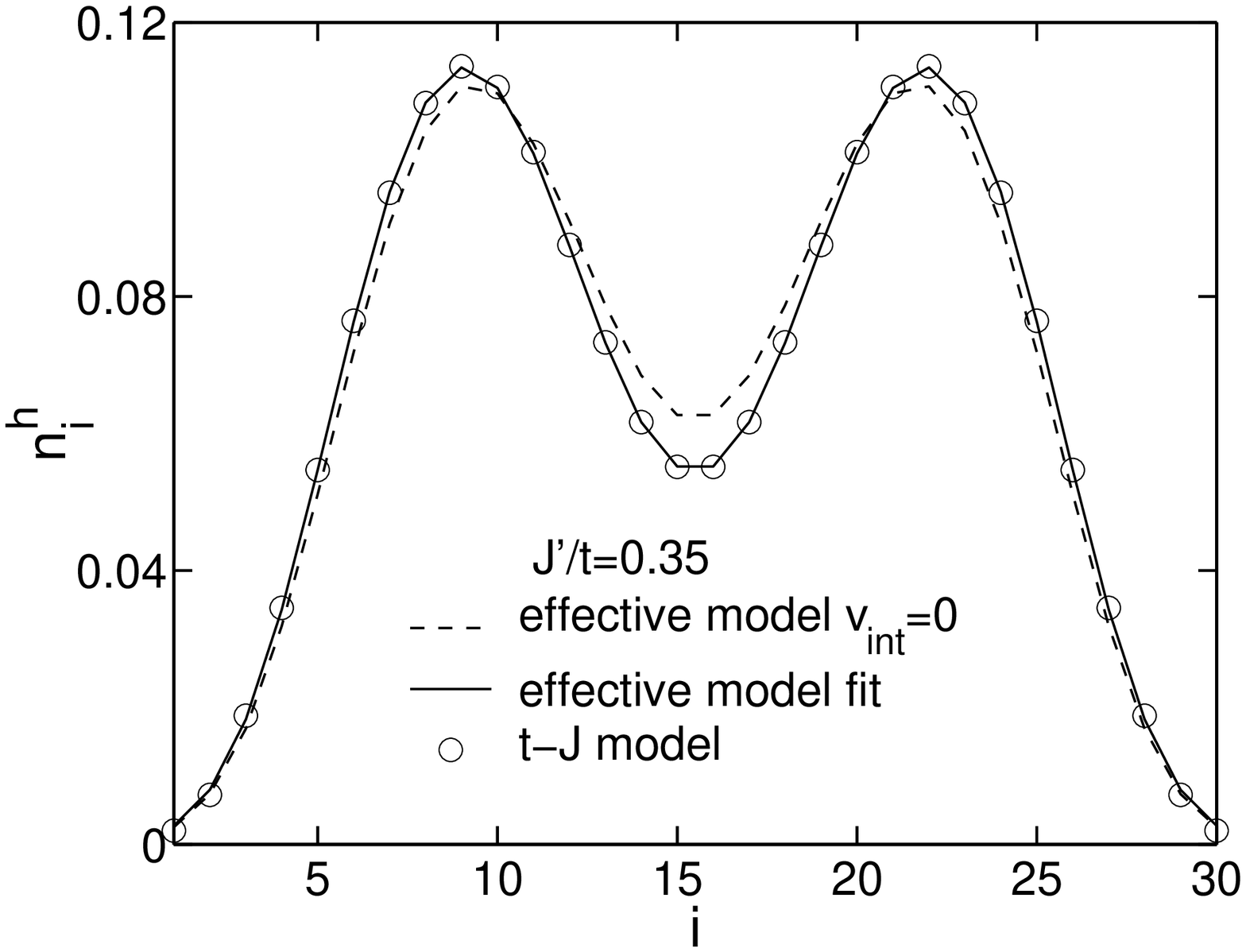}\\
\epsfxsize=\linewidth
\epsffile{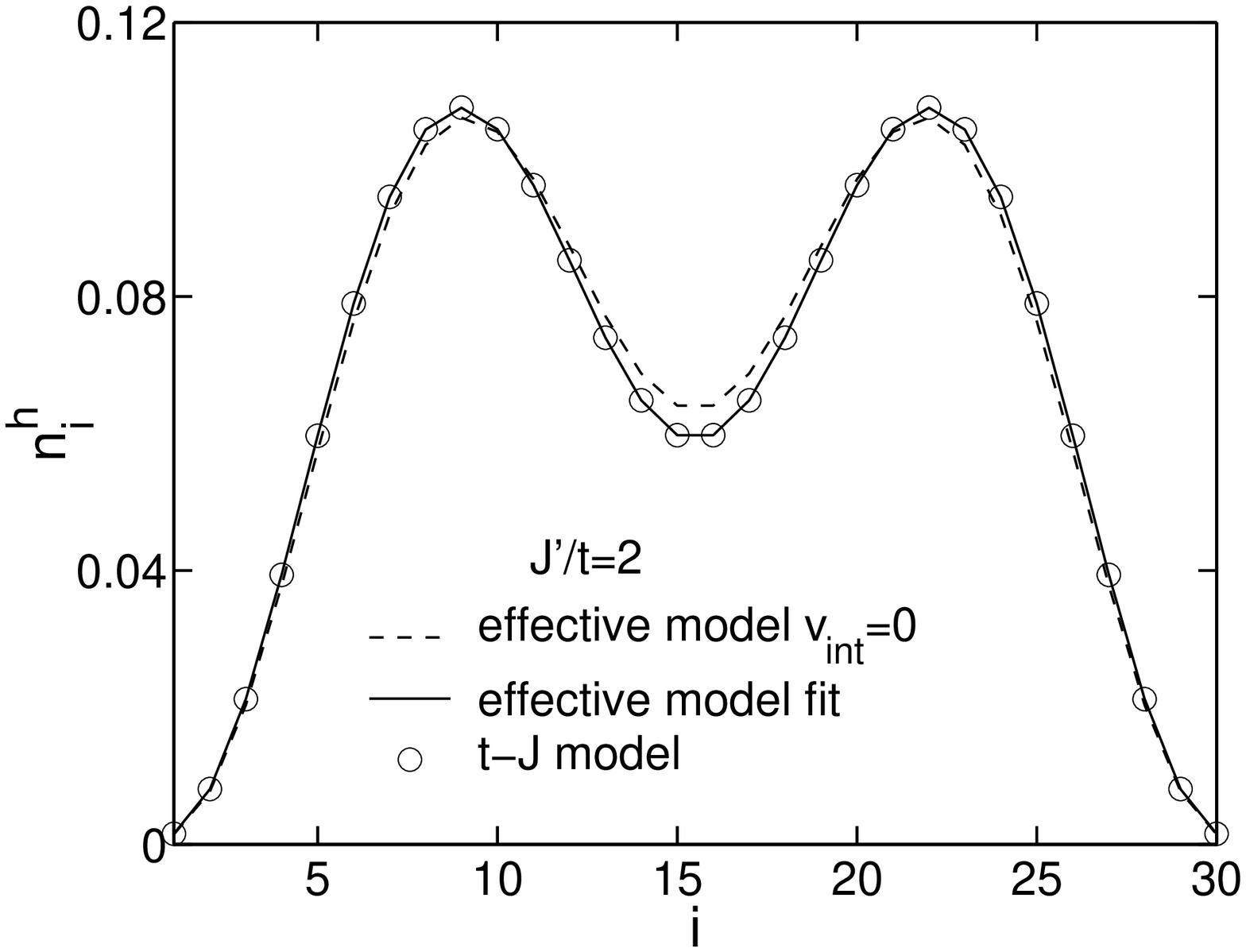}
\caption
{
  Hole density $n_i^h$ for $J'=0.35\, t$ and $J'=2\, t$ calculated for four 
  holes on a ($30\times 2$) \tj ladder computed directly and with the
  effective model ($J = 0.35\, t$, $t = t'$).
  For the effective model the data for $v_{\rm int}=0$ and the best fit
  is shown.
} 
\label{intfitfig1}
\end{center}
\end{figure}

\begin{figure}
\begin{center}
\epsfxsize=\linewidth
\epsffile{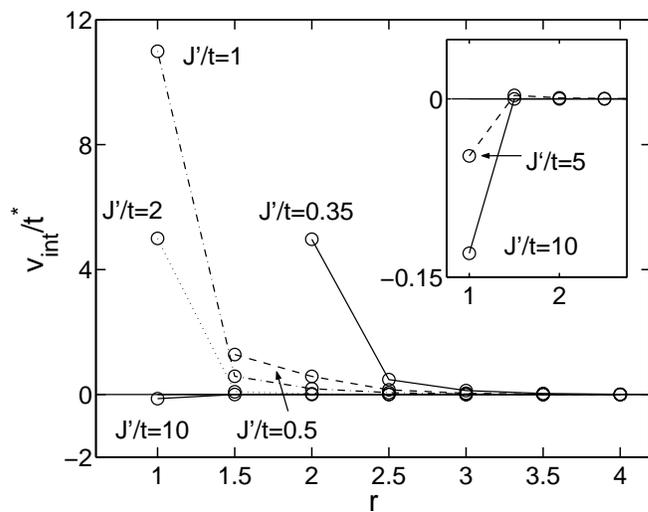}
\caption
{
  Interaction parameter $v_{{\rm int}}(r\geq r_{\rm min})$ between
  the hcb obtained from ($30\times 2$) \tj ladders for various
  $J'$ ($J = 0.35\, t$, $t = t'$).
} 
\label{vintplot}
\end{center}
\end{figure}

Since the hole pair becomes broader with decreasing $J'$,
it was useful to let $r_{min}$ vary. 
Again, we have made fits between the density profiles as shown
in Fig.~\ref{intfitfig1} and obtained the 
parameters represented in Table~\ref{tab:fitpar}.
Figure~\ref{vintplot} shows $v_{\rm int}$ for different $J'$ values.
Starting from the isotropic case, the interaction is repulsive and
becomes less long ranged as $J'$ increases. For $J' = 5\,t$, $v_{\rm int}$
becomes negative for $r=2$. The same holds for $J'= 10\,t$, where 
$v_{\rm int}$ vanishes for $r>2$. This can be expected from a perturbative
approach as can be found in Ref.~\onlinecite{troyer}, where the hole pairs on
a \tJ ladder have been mapped to hcb on a chain with next-nearest
neighbor interaction. 
%Note that $V_b(1)=v_{\rm int}(r=1)=-0.130\,t$.

\begin{figure}
\begin{center}
\epsfxsize=\linewidth
\epsffile{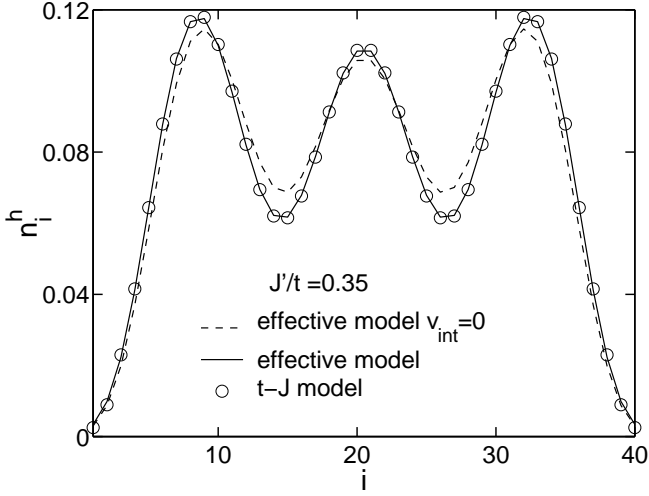}\\
\epsfxsize=\linewidth
\epsffile{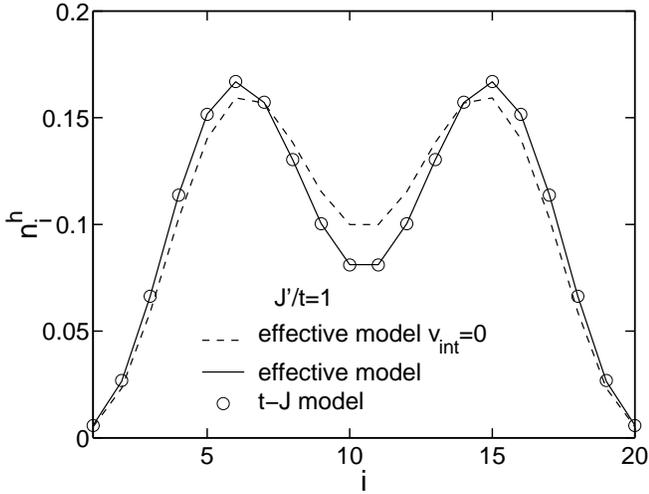}
\caption
{
  Hole density $n_i^h$ for $J' = 0.35\, t$ and $J'= t$ for six holes 
  on ($40\times 2$) and four holes on ($20\times 2$) sites respectively,
  computedwith the interaction potentials obtained from fits with two
  hole-pairs on $30$ sites ($J = 0.35\, t$, $t = t'$). For the effective
  model the data for $V_{\rm int}=0$ is also shown.
  } 
\label{intfitfig2}
\end{center}
\end{figure}

We have tested the results obtained in this way by 
comparing the density profiles for
various numbers of hole pairs and system lengths. From strong coupling
down to the isotropic case we find good agreement between the density
profiles obtained from the \tj model and the effective model, as can be
seen in Fig.~\ref{intfitfig2}.

We should note here that the fits for $v_{\rm int}$ were not as stable as
the fits for $v_{b}$. Varying simultaneously the parameters $v_1$, $v$
and $\xi$ in Eq.~(\ref{vb}), one can find another set of parameters
which also gives a good fit. However, we found that the Luttinger liquid
parameter $K_\rho$, which we calculate in the next section,
only weakly depends on this ambiguity. Using only two parameters to
determine $v_{\rm int}$ suppressing $v_1$ in Eq.~(\ref{vb}), one can also
fit the density profiles with a different set of parameters, but in
this case $K_\rho$ is strongly affected. 

%
%------------------------------------------------------------------------------
% The hopping matrix element $t^*$
%------------------------------------------------------------------------------
%
\subsection{The hopping matrix element $t^*$}

Up to now the hopping matrix element $t^*$ did not play any role.
We were only interested in the hcb density profile, which is not
affected by the energy scale fixed by $t^*$.

\begin{figure}
\begin{center}
\epsfxsize=\linewidth
\epsffile{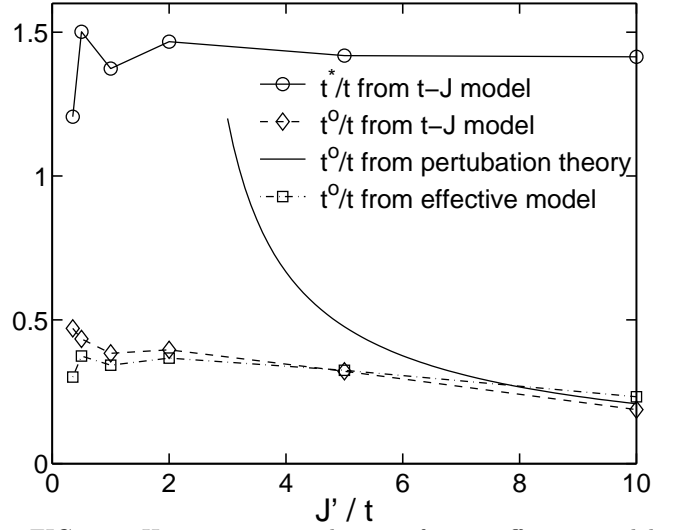}
\caption{
  Hopping matrix elements for our effective model and for the model
  in Ref.~\protect\onlinecite{troyer} obtained for the large $J'$ limit
  ($J = 0.35\, t$, $t = t'$). Explanations are in the text. 
} 
\label{tstar}
\end{center}
\end{figure}

We compute $t^*$ by finite size scaling with the ansatz
that the difference between the ground state energy for
two holes $E_{\rm 2h}$ and for zero holes $E_{\rm 0h}$ in
the \tj model is given by 
\ba\label{fssc_em}
E_{\rm 2h}(L)-E_{\rm 0h}(L) &\simeq& {\rm const}+
  t^* E_{\rm eff}(L)\ .
\ea
Here $E_{\rm eff}$ denotes the ground state energy of the
corresponding effective model with one boson and with $t^*=1$.
We obtained $t^*$ by fitting
Eq.~(\ref{fssc_em}) to a straight line. The result is shown in
Fig.~\ref{tstar}. 

According to the effective model given in Ref.~\onlinecite{troyer}, one can 
obtain the corresponding effective hopping matrix element $t^o$ by a
finite size scaling with the ansatz
\ba\label{fssc_em_tr}
E_{\rm 2h}(L)-E_{\rm 0h}(L) &\simeq& {\rm const}+ t^o (\pi/(L+1))^2 \ .
\ea
Here, $E_{\rm 2h}(L)$ and $E_{\rm 0h}(L)$ are for the \tj
model.
A perturbative estimate for $t^o$ is given in Ref.~\onlinecite{troyer}.
For large $J'$ one obtains
\ba\label{tstar_pert}
t^o &=& 2\,t^2/ (J'-4\,t^2/J') \ .
\ea
Figure~\ref{tstar} shows $t^o$ versus $J'$. 
It can be seen that for $J'\to \infty$ the $t^o$ obtained 
by perturbation theory tends to the one obtained by finite
size scaling.
Note that $t^o$ stands for the hopping of a hole pair from
one rung to another, whereas in our effective model $t^*$
mediates between $i$ and $i+\frac{1}{2}$.
We can obtain $t^o$ from $t^*$ by expanding the
single boson dispersion of the effective model
for small momenta $k$.
In this way we obtain
\ba
t^o&=&\frac{{t^*}^2}{\sqrt{\epsilon^2+(4\,t^*)^2}} \ .
\ea
Figure~\ref{tstar} shows that $t^o$ obtained in this way from
our ef\-fective model fits very well to the one obtained by
finite size scaling.

\begin{figure}
\begin{center}
\epsfxsize=\linewidth
\epsffile{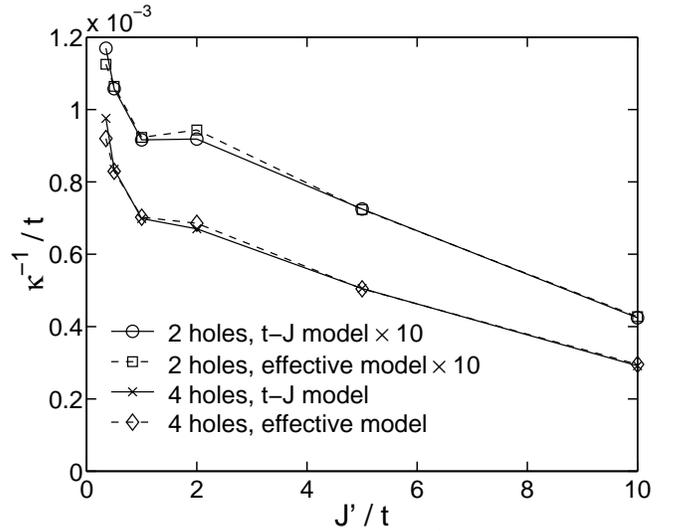}
\caption
{
  Inverse compressibility $\kappa^{-1}$ for the \tj Ladder
  computed directly and with the corresponding effective
  model for 2 holes (magnified $10\times$) and 4 holes for
  a $40 \times 2$ \tj ladder ($J = 0.35\, t$, $t = t'$).
} 
\label{kappa}
\end{center}
\end{figure}

An independent test for $t^*$ comes from the comparison of
the inverse compressibility $\kappa^{-1}$ obtained from the
\tj model and the effective model. The inverse compressibility
could be obtained from
\ba
\kappa^{-1}(\rho)
&=&
\frac{\rho^2}{\Omega}\,\frac{\partial^2 E_0(\rho)}{\partial \rho^2} \ .
\ea
Here the volume $\Omega=2\,L$ for the \tJ model and $\Omega= 2\,L-1$
for the effective model. With $N$ we denote the hole or hcb number and
with $\rho=N/\Omega$ the corresponding density.
The second derivative of the ground state energy with respect to the
particle number $N$ was calculated by the discretization
\ba\label{eq:secderiv}
\frac{ \partial^2 E_0}{\partial N^2}
&\simeq& 
\frac{E_0(N+\Delta)-2\,E_0(N)+E_0(N-\Delta)}{\Delta^2} \ .
\ea
Here $\Delta=2$ and $\Delta=1$ has been used for
the \tj model and the effective model respectively.
Figure~\ref{kappa} shows the results for $\kappa^{-1}$
computed for a ($40 \times 2$) \tj ladder with 2 holes
and 4 holes.
As can be expected, the results agree better for large
$J'$, where the hole pairs are narrower and the
interaction between pairs is less long-ranged, but the
agreement is satisfactory for all values of $J'/t$.

%
%------------------------------------------------------------------------------
% Section IV : Luttinger liquid parameter
%------------------------------------------------------------------------------
%
\section{Luttinger liquid parameter $K_{\rho}$}

In the previous section we obtained the interaction between two hcb
in the effective model (i.e. two hole pairs in the \tj model).
In order to obtain information about the long range correlations we
have calculated the Luttinger liquid parameter $K_{\rho}$ in the effective
model.

Our effective model belongs to the same universality class as the 
hcb model with next-nearest neighbor interaction, which in turn is equivalent
to the XXZ model and which has been solved exactly by a bosonization
approach and conformal field theory.\cite{luther_peschel,haldane}
From Ref.~\onlinecite{luther_peschel} we obtain a power law decay at large
distances for the charge density wave correlations and the superconducting
correlations given as
\ba
\langle N_r N_0 \rangle &\sim&
{\rm const}\times r^{-2} + {\rm const}\times \cos(2\pi\rho\, r)\, r^{-2\,K_{\rho}}
\label{cdw}
\\
\langle B_r^{\dag} B_0 \rangle &\sim&
{\rm const}\times r^{-\frac{1}{2\,K_{\rho}}}\ \ .
\label{sccorr}
\ea
Here $\rho$ denotes the electron density on the ladder and 
half filling corresponds to $\rho=1$.
These relations show that the superconducting
correlations $\langle B_r^{\dag} B_0 \rangle$ are
dominant if $K_{\rho}>\frac{1}{2}$.

For hcb in one dimension, $K_{\rho}$ can also
be obtained from the relations \cite{schulz}
\ba
K_{\rho} &=& \pi v_c L \biggl( \frac{ \partial^2 E_0}
{\partial N_b^2}\biggr)^{-1}\\ 
v_c K_{\rho} &=& \frac{\pi}{L}\,\frac{\partial^2 E_0(\Phi)}
{\partial \Phi^2}\bigg|_{\Phi=0} \ \ .\nn
\ea
Here $E_0$ denotes the ground state energy for a closed
ring of length $L$ with $N_b$ hcb and
$E_0(\Phi)$ is the ground state energy of the system
penetrated by a magnetic flux $\Phi$ which modifies the
hopping by the usual Peierls phase factor, $ t \mapsto t
\exp(\pm\, i \Phi/L)$. From these two equations the charge
velocity $v_c$ can be eliminated.
The second derivative of the ground state energy with
respect to the particle number was calculated by the
discretization given in Eq.~(\ref{eq:secderiv}) with
$\Delta=1$.
We used exact diagonalization for system lengths between 
$32$ and $220$ and with $N_b$ between 2 and 4.
Analogously, the second derivative with respect to the
magnetic flux was calculated for the same configurations.

\begin{figure}
\begin{center}
\epsfxsize=\linewidth
\epsffile{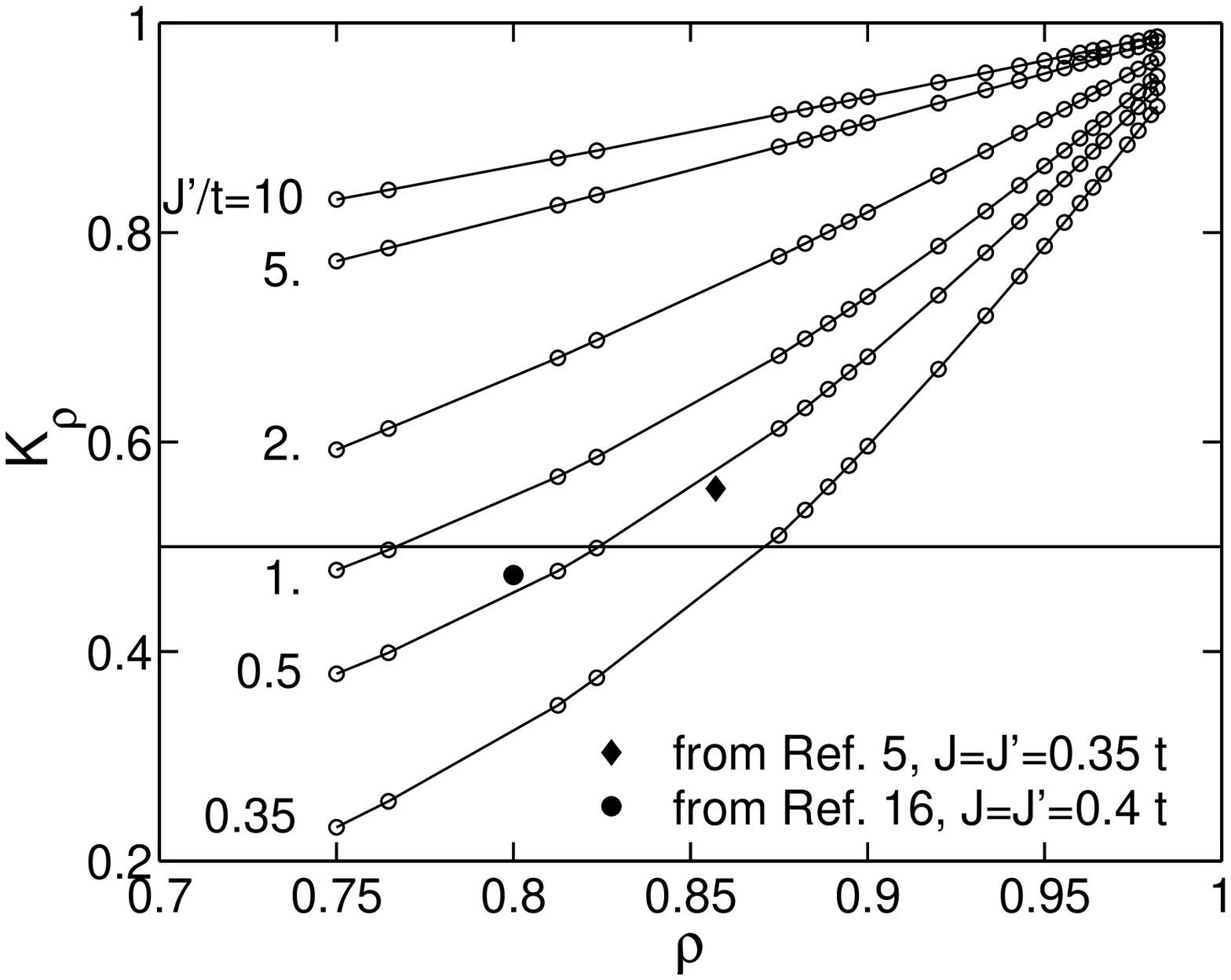}
\caption
{
  $K_{\rho}$ as a function of $\rho$ for various $J'$
  (half filling is given by $\rho=1$, $J = 0.35\, t$, $t = t'$).
}
\label{krho_rho}
\end{center}
\end{figure}

\begin{figure}
\begin{center}
\epsfxsize=\linewidth
\epsffile{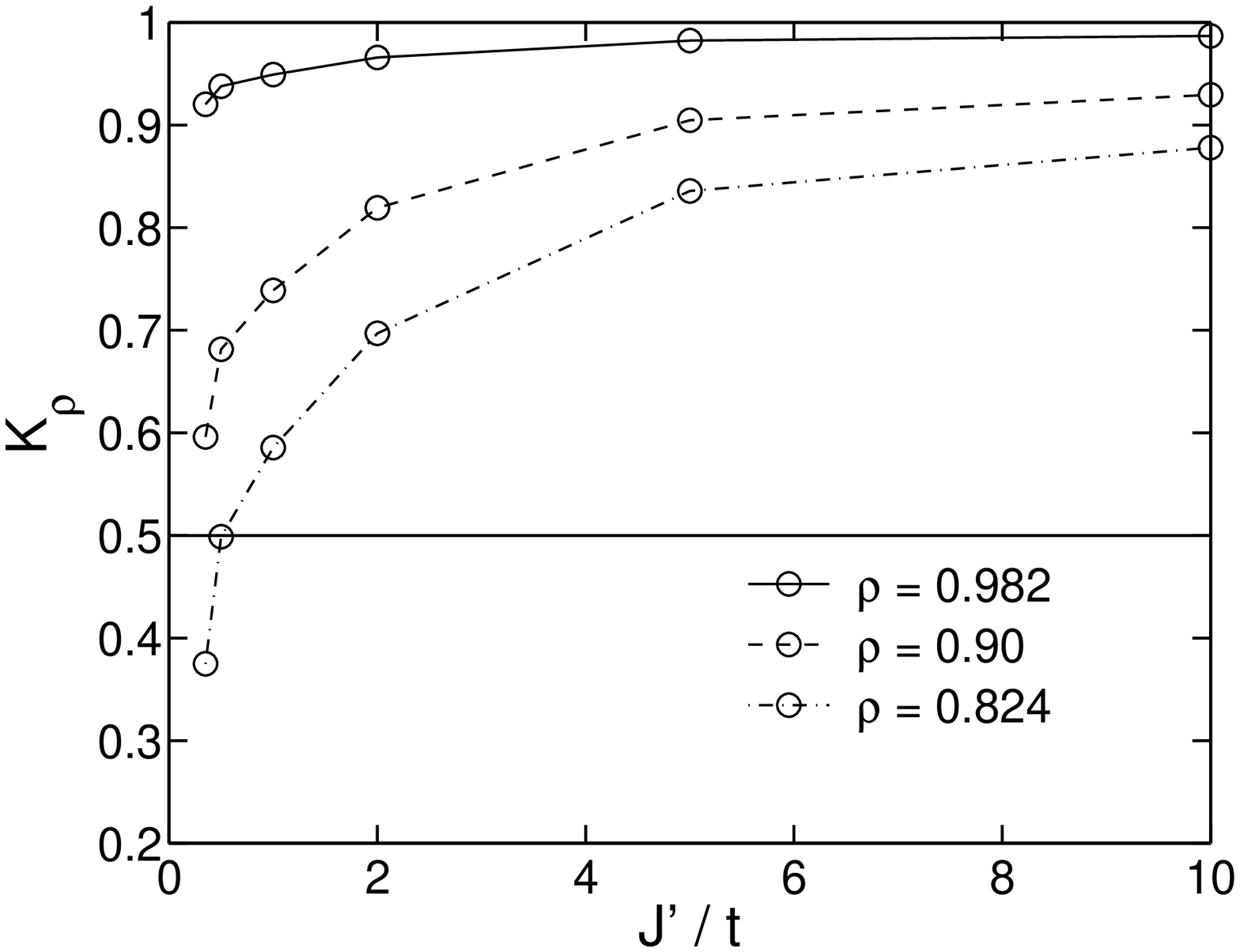}
\caption
{
  $K_{\rho}$ as a function of $J'$ for various electron
  densities $\rho$ (half filling is given by $\rho=1$,
  $J = 0.35\, t$, $t = t'$).
}
\label{krho_jp}
\end{center}
\end{figure}

We calculated the Luttinger liquid parameter $K_{\rho}$
for the interaction potentials given by the parameters
in Table~\ref{tab:fitpar} and found the universal value
$K_{\rho}=1$ as half filling is approached, for any
value of $J'$ between strong coupling and the isotropic
case (Figs.~\ref{krho_rho} and \ref{krho_jp}). For $N_b/L \to 0$,
corresponding to a very dilute hcb gas, we have
$K_{\rho}=1+O(N_b/L)$, independent of the value of the
interaction and consistent with Refs.~\onlinecite{schulz_99,poilblanc}.
Down to $\rho \simeq 0.875$ the superconducting 
correlations are dominant,
since $K_{\rho}> \frac{1}{2}$ and since we are far away 
from phase separation.\cite{troyer}
Below $\rho \simeq 0.875$, $K_{\rho}$ becomes less than 
one half and at  one quarter doping ($\rho = 0.75$)
$K_{\rho} < \frac{1}{2}$ for $J'\leq t$.
Figure~\ref{krho_rho} shows also earlier results on small clusters from 
Ref.~\onlinecite{troyer} and Ref.~\onlinecite{hayward},
obtained using exact diagonalization. The deviations from our results are
most probably due to finite size effects close to half filling in
Ref.~\onlinecite{troyer} and Ref.~\onlinecite{hayward}.

\begin{figure}
\begin{center}
\epsfxsize=\linewidth
\epsffile{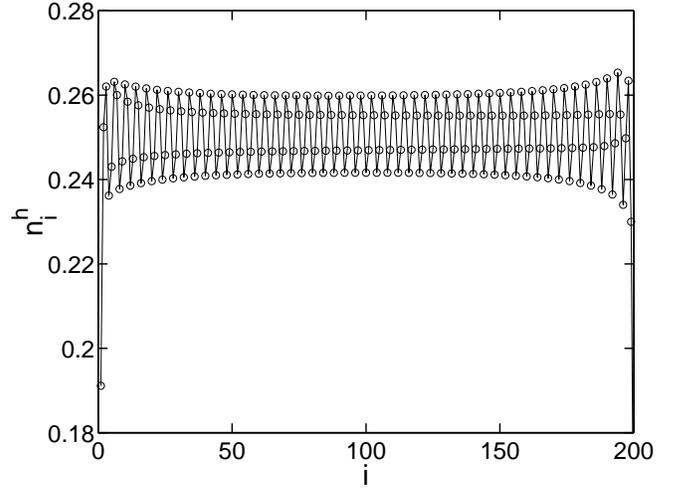}
\caption
{
  Density profile $n_i^h$ computed with the effective model for
  a ($200\times 2$) \tj ladder at one quarter doping ($\rho=0.75$)
  with the parameters of the isotropic case ($J = J' = 0.35\, t$, $t = t'$),
  showing a pinned CDW.  
} 
\label{fig:cdwem}
\end{center}
\end{figure}

\begin{figure}
\begin{center}
\epsfxsize=\linewidth
\epsffile{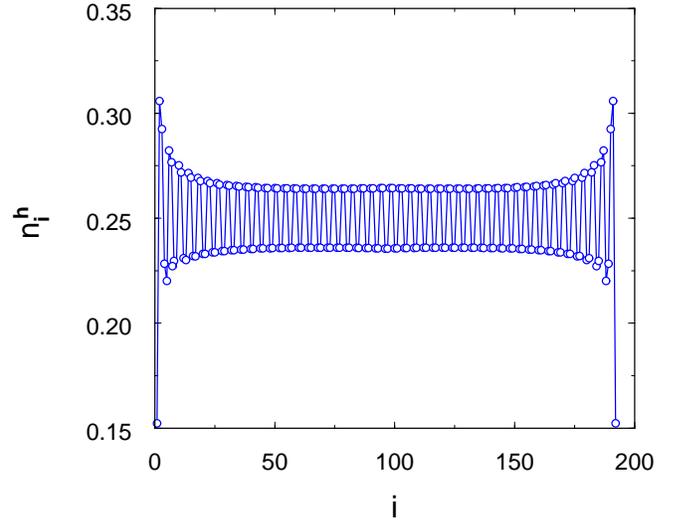}
\caption
{
Hole density for a ($192 \times 2$) \tj ladder at one quarter
doping ($\rho=0.75$) showing a pinned CDW. Here $J'=J=0.3\,t$ and $t'=t$.
} 
\label{fig:cdwtj}
\end{center}
\end{figure}

We briefly want to discuss the possibility of commensurability effects.
At a filling of $\rho = 0.75$, one might expect that commensurability
effects would stabilize long ranged charge density wave ordering,
since $K_\rho$ is quite small ($K_\rho = 0.232$ for the isotropic case).
In the finite, open systems studied here, a static CDW is pinned by
the boundaries, as can be seen in Fig.~\ref{fig:cdwtj} for the \tj model
as well as in Fig.~\ref{fig:cdwem} for the effective model.
A careful analysis using large systems and finite size scaling is
necessary to determine if the CDW order is long ranged.
The decay of CDW oscillations away from the edges of the system is
quite slow and is consistent with long ranged CDW order, but a very
slow algebraic decay cannot be ruled out. 
A more detailed analysis of the effective model (using 200, 240, 280, 320
and 400 sites) shows a decreasing charge gap with increasing system size.
Finite size scaling is consistent with a vanishing or possibly
a small charge gap in the infinite system.

%
%------------------------------------------------------------------------------
% Conclusions
%------------------------------------------------------------------------------
%

\section{conclusions}

The effective model derived in this paper works well for
the hole density of a two-leg \tj ladder for 
various fillings. In other words, the low energy physics
of hole pairs on a ladder can be well described by a
model of hard core bosons on a chain with each boson
representing a pair of holes.
The interaction between the hard core bosons was determined
by fitting the density profile obtained with the effective
model to that of the \tj model, taking
into account the inner structure of the hole pair given
by the hole-hole correlation function. Starting from the
isotropic case, with equal exchange couplings on the rungs 
and legs the interaction between two hole pairs
is long ranged and repulsive but becomes attractive and
of nearest neighbor type when the strong coupling regime 
is approached. The same holds for the interaction of
a hole pair with the boundaries.
We choose a simple form for the interaction between the
bosons in order to use only a few parameters.
The results obtained from the effective model
are insensitive to the specific ansatz used for
this interaction. 
The Luttinger liquid parameter $K_{\rho}$ has been calculated for 
electron densities from $\rho=0.982$ down to $\rho=0.75$
(half filling corresponds to $\rho=1$). Down to 
$\rho \simeq 0.875$ the superconducting correlations are
found to be dominant and $K_\rho \to 1$ for $\rho \to 1$.
For commensurate filling, $\rho=0.75$, there might be true
charge density wave ordering and a charge gap.
Further investigations are necessary to clarify this question.
The hopping matrix element for the bosons in the effective model,
which allows one to calculate the inverse compressibility, could 
also be determined. 
Comparing to the \tj model we find good agreement for strong rung
couplings and only small deviation (less than $\pm 10\%$) near
the isotropic case.

\begin{figure}
\begin{center}
\epsfxsize=\linewidth
\epsffile{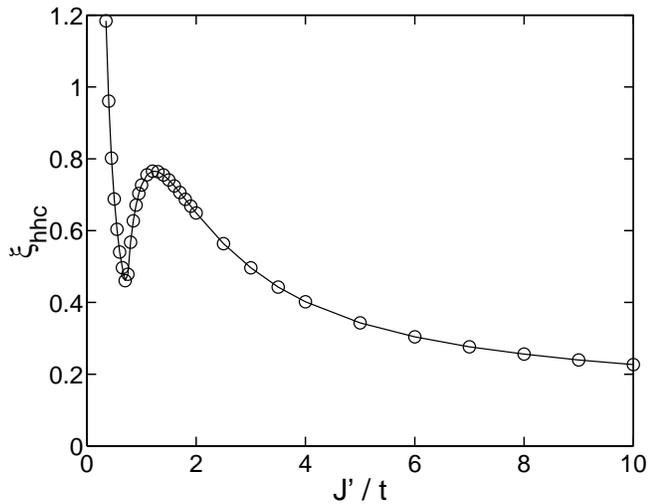}
\caption{Correlation length $\xi_{hhc}$ of the hhc-function $g(r)$ as a
  function of $J'$ ($J = 0.35\, t$, $t = t'$) measured on a
  ($40\times 2$) \tj ladder.}
\label{xicorr}
\end{center}
\end{figure}

An interesting feature appeared in the hole-hole correlation
function. The correlation length, $\xi_{hhc}$, does not
monotonically increase as one approaches the isotropic case 
from strong rung coupling. Instead, it decreases in the
interval $0.7\,t\leq J' \leq 1.2\,t$.
This unexpected behavior can be traced to the fact
that the interaction between holes is dominated by
the simple rung exchange bonds in the strong rung
coupling regime and by the diagonal rung exchange
bonds near the isotropic case, $J'\to J$. 
Reflections of the structure in $\xi_{hhc}$ as a
function of $J'$, shown in
Fig.~\ref{xicorr}, can also be found in other properties,
e.g. the interaction of the effective bosons 
with the boundaries and the compressibility.

\acknowledgements

We wish to thank Karyn Le Hur for helpful discussions.
SRW acknowledges support from the NSF under grant \#DMR98-70930.
The DMRG calculations have been performed on the SGI Cray SV1
of ETH Z\"urich. MT was supported by the Swiss National Science
Foundation.

%
%------------------------------------------------------------------------------
% References
%------------------------------------------------------------------------------
%

\newpage

%fig01

%fig02

%fig03

%fig04

%fig05

%fig06

%fig07

%fig08

%fig09

%fig10

%fig11

%fig12

%fig13

%fig14

%fig15

%fig16

\begin{table}
\begin{center}
\caption{
  Parameters for the interaction potential $V_{\rm int}$ given in
  Eq.~(\ref{interaction}) obtained from the density profiles 
  of a ($30\times 2$) \tj ladder for various $J'$ ($J = 0.35\, t$, $t = t'$).
}\label{tab:fitpar}
\end{center}
\begin{tabular}{r c r r r}
$J'$& $r_{min}$ & $v_1$ & $v$ & $\xi$ \\
\hline
0.35 & 2 & 4.973  & 0.478  & 0.376 \\
0.5  & $\frac{3}{2}$ & 1.28   & 0.580  & 0.373 \\
1.   & 1 & 10.99  & 0.586  & 0.439 \\
2.   & 1 & 5.00   & 0.089  & 0.363 \\
5.   & 1 & -0.048 & 0.0029 & 0.400 \\
10.  & 1 & -0.13  & 0.0    & -\\
\end{tabular}
\end{table}

\end{document}